\documentclass[12pt]{article}
\usepackage{amsmath}
\usepackage{graphicx}
\usepackage{enumerate}

\usepackage[utf8]{inputenc}
\usepackage{amsmath,amsthm,amssymb, graphicx, multicol, array, appendix, hyperref, cleveref, enumitem, verbatim, multirow, comment}
\usepackage[table,xcdraw]{xcolor}
\usepackage{natbib, setspace, caption, subcaption, algpseudocode, algorithm}

 \newtheorem{example}{Example}

\def\hZ{\hat Z}
\def\hK{\hat K}

\def\hP{\hat P}
\def\hG{\hat G}
\def\hH{\hat H}
\def\hU{\hat U}
\def\hS{\hat S}
\def\tr{\operatorname{Tr}}
\def\hT{\hat \Theta}
\def\hz{\hat z}

\def\v1{\mathbf{1}}
\def\Arc{\operatorname{acosh}}

\newtheorem{theorem}{Theorem}[section]

\newtheorem{lemma}[theorem]{Lemma}

\theoremstyle{remark}
\newtheorem{remark}{Remark}[section]
\newtheorem{property}[theorem]{Property}

\usepackage{url} 

\newcommand{\blind}{1}

\begin{document}

\def\spacingset#1{\renewcommand{\baselinestretch}%
{#1}\small\normalsize}

\spacingset{1}


\if1\blind
{
  \title{\bf Hyperbolic Network Latent Space Model with Learnable Curvature}
  \author{Jinming Li, Gongjun Xu, and Ji Zhu
\hspace{.2cm}\\
    Department of Statistics, University of Michigan\\
}
  \maketitle
} \fi

\if0\blind
{
  \bigskip
  \bigskip
  \bigskip
  \begin{center}
    {\LARGE\bf Hyperbolic Network Latent Space Model with Learnable Curvature}
\end{center}
  \medskip
} \fi

\bigskip
\begin{abstract}
Network data is ubiquitous in various scientific disciplines, including sociology, economics, and neuroscience. Latent space models are often employed in network data analysis, but the geometric effect of latent space curvature remains a significant, unresolved issue. In this work, we propose a hyperbolic network latent space model with a learnable curvature parameter.
We theoretically justify that learning the optimal curvature is essential to minimizing the embedding error across all hyperbolic embedding methods beyond network latent space models. 
A maximum-likelihood estimation strategy, employing manifold gradient optimization, is developed, and we establish the consistency and convergence rates for the maximum-likelihood estimators, both of which are technically challenging due to the non-linearity and non-convexity of the hyperbolic distance metric. We further demonstrate the geometric effect of latent space curvature and the superior performance of the proposed model through extensive simulation studies and an application using a Facebook friendship network.
\end{abstract}

\noindent%
\vfill

 \newpage
\spacingset{1.85} 

\section{Introduction}

With recent advances in science and technology, network data that contain relational information among observations are becoming more prevalent than ever. Networks emerge ubiquitously in many fields including social media \citep{traud2012social}, brain science \citep{bullmore2009complex}, and economics \citep{chaney2014network}. Statistical models are especially useful for understanding the network structure and characterizing the generative process of edges. Contrasting classical scenarios where observations are assumed to be independent, network data often display higher-order, complex dependency structures such as transitivity and modularity \citep{newman2018networks}. 

Latent space models, recognized for their flexibility in modeling complex network structures, are widely used in network analysis \citep{hoff2002latent, ma2020universal,zhang2022joint} and have demonstrated their power in capturing various features commonly observed in real-world networks \citep{ward2007persistent, ward2007disputes, ward2011network, friel2016interlocking}.
Under the latent space models, each node is assigned a latent position $z_i$, and the network edges are generated conditioning on these latent positions. 
One popular class of network latent space models is the distance model, first proposed in \cite{hoff2002latent}. It assumes that the linkage probability of two nodes $i$ and $j$ is
$\sigma(d(z_i, z_j)) \in [0, 1]$, where $d(z_i,z_j)$ is the distance between nodes $i$ and $j$ in the latent space. The link function $\sigma(x)$ is strictly decreasing so that nodes closer in the latent space are more likely to be connected. The model has nice properties that it naturally admits transitivity and homophily due to the distance triangular inequality: given nodes A, B, and C, if A and B, A and C are close in the latent space respectively, then B and C are close as well. In \cite{hoff2002latent}, the authors set $\sigma(x) = \operatorname{logistic}(\alpha -x)$ and $d(z_i,z_j) = |z_i-z_j|$ where $z_i \in \mathbb{R}$ and $\alpha$ controls the maximal linking probability. \cite{ma2020universal} further studied an inner product latent space model, which in a special case can be formulated as a low-dimensional Euclidean distance model with $\sigma(x) = \operatorname{logistic}(\alpha -x^2/2)$ and $d(z_i,z_j)=\|z_i-z_j\|$. 

While the Euclidean space is traditionally employed as the default latent space in existing literature, its suitability for modeling network data remains an open question. In recent years, there has been accumulating evidence suggesting that hyperbolic space, a metric space characterized by constant negative curvature, may be more appropriate for network data modeling. For instance, \cite{krioukov2010hyperbolic} proposed a latent space model with hyperbolic distance metric and showed that the node degree distribution naturally follows the power-law, a common characteristic of numerous real-world social and internet networks. Further, \cite{kennedy2016hyperbolicity} defined an analogy for curvature in discrete graphs, termed ``hyperbolicity", and used it to show that hyperbolic geometry prevails in social networks, internet networks, and collaboration networks. In the field of machine learning, researchers have proposed graph embedding models with hyperbolic geometry, reporting superior performances over their Euclidean counterparts \citep{nickel2017poincare, nickel2018learning, chami2019hyperbolic}. 

The focus on latent space geometry in statistical modeling and inference has gained increased attention in the statistics community. \cite{smith2019geometry} conducted extensive simulation studies comparing the global and local characteristics of networks generated from distance models across spherical, Euclidean, and hyperbolic geometries. Their findings suggest that networks generated from hyperbolic models are more similar to networks collected from various applications than those produced from other geometries. Recently, \cite{lubold2020identifying} considered the problem of conducting statistical inference of the latent space curvature and proposed a model-based hypothesis testing framework. The testing procedure, motivated by the connection between the eigenvalues of the distance matrix and latent space curvature, can be used to decide whether the latent space geometry is  Euclidean, hyperbolic, or spherical. 
 
However, the geometric impact of curvature on latent space modeling remains insufficiently understood in the existing literature. Motivated by the fact that Euclidean and hyperbolic spaces with different curvatures are not isometric, regardless of whether they share the same dimension (see Property \ref{property:impossibility_of_isometry} for more details), this work aims to integrate the latent space curvature into the model as a learnable parameter. Moreover, statistical inference problems related to the latent space geometry or latent positions in hyperbolic spaces have been largely underexplored in the literature. This gap motivates us to develop estimation algorithms and to establish theoretical foundations and inference procedures within the maximum likelihood framework. Our main contributions are summarized as follows.

First, from the modeling perspective, we propose a unified hyperbolic latent space modeling framework with a learnable curvature parameter $- K < 0$, which allows us to embed networks in hyperbolic spaces with different curvatures. In particular, the model can be viewed as a generalization of the Euclidean distance model in that the hyperbolic distance becomes the Euclidean distance when $K \to 0$. Most existing hyperbolic models, e.g. \citet{smith2019geometry} and \citet{nickel2017poincare}, however, only considered a default curvature of $-1$ and did not recognize the geometric effect of the curvature. We further investigate the identifiability of latent space curvature and address the unanswered question about the interplay between the effect of geometry and the distribution of latent positions \citep{smith2019geometry}, demonstrating that the geometric effect of curvature is unique and cannot be substituted by the distribution of latent positions.

Second, from the theoretical perspective, we establish consistency and convergence rates for the maximum-likelihood estimators of the curvature parameter and latent positions. Compared to existing literature, while \cite{shalizi2017consistency} showed that maximum-likelihood estimators of hyperbolic distance models are consistent, their result holds only when the latent space curvature is known and requires some strong asymptotic assumptions on the network linkage probabilities. Moreover, the analytic approach used in \cite{shalizi2017consistency} cannot be used to derive the consistency rates of the estimators. Our analysis adopts totally different techniques from hyperbolic geometry and low-rank matrix completion, which to the best of our knowledge establish the first consistency rate result. Notably, our theorem shows that embedding error is inevitable when the latent space curvature is misspecified, which highlights the geometric effect of curvature for all hyperbolic embedding methods beyond network latent space models. 

Last, from the computational perspective, we develop a manifold gradient descent algorithm as well as initialization methods to overcome the challenges in non-convex optimization of the likelihood function. We also conduct extensive simulation studies to illustrate the impact of latent space curvature and analyze a Facebook friendship network with the proposed model, demonstrating its interpretable visualization and great performance.


The rest of this paper is organized as follows. In Section 2 we introduce the basics of hyperbolic geometry and propose the hyperbolic latent space model for network data. Section 3 describes the estimation procedure, and the main theoretical results are presented in Section 4. 
To evaluate the performance and effectiveness of the proposed model, we conduct several simulation studies, the results of which are reported in Section 5,
where we also propose a bootstrap-based procedure for statistical inference of the curvature parameter. 
In Section 6, we apply the hyperbolic latent space model to analyze a real-world Facebook friendship network, demonstrating the practical applicability of the proposed approach. Furthermore, in Section 7, we discuss potential future directions and extensions of the model.
Proofs of the theoretical results are presented in the Supplementary Materials.

\section{The Hyperbolic Network Latent Space Model}

\subsection{Preliminary on Hyperbolic Geometry}

Hyperbolic spaces are metric spaces with a constant negative sectional curvature, while Euclidean spaces have a curvature of constant 0. For hyperbolic spaces to have a well-defined notion of curvature, they must have dimensions greater than one. Intuitively, the negative curvature of hyperbolic spaces allows for more expansive room compared to Euclidean spaces, enabling them to accommodate more heterogeneous data. Specifically, the area of a hyperbolic disk grows exponentially with its radius, whereas in Euclidean space, the growth rate is quadratic. For a more detailed comparison of the basic geometric properties of hyperbolic, Euclidean, and spherical spaces, refer to \cite{krioukov2010hyperbolic}.

\begin{figure}[ht]
\centering
\begin{subfigure}{.5\textwidth}
  \centering
  \includegraphics[width=.5\linewidth]{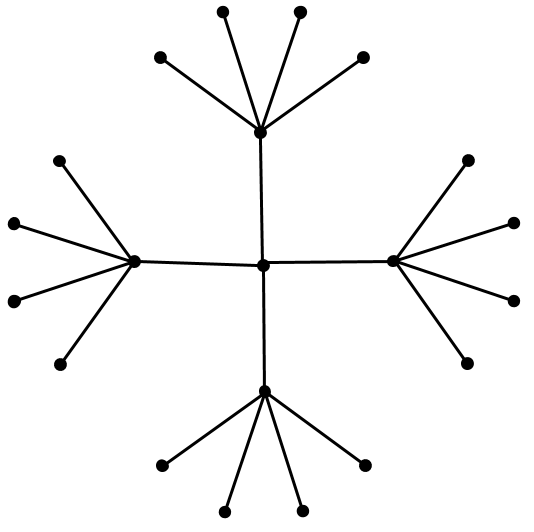}
  \label{fig:eu_embed_tree}
\end{subfigure}%
\begin{subfigure}{.5\textwidth}
  \centering
  \includegraphics[width=.5\linewidth]{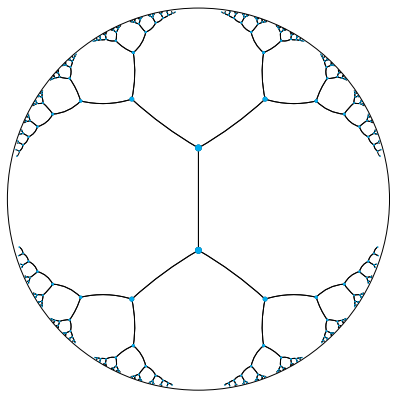}
  \label{fig:hy_embed_tree}
\end{subfigure}
\caption{An example of embedding tree-structured data. Left: embedded with $2$-dimensional Euclidean space; Right: embedded with $2$-dimensional Hyperbolic space.}
\label{fig:embed_tree}
\end{figure}

One motivating example that helps illustrate the difference between hyperbolic space and Euclidean space is the embedding of trees, which are a special type of graphical data with a hierarchical node structure. \Cref{fig:embed_tree} illustrates an example of attempting to isometrically embed a $4$-ary tree into $2$-dimensional Euclidean and hyperbolic spaces. As shown in the left plot, when the depth of the tree is only 2, the leaf nodes are quite close to each other, resulting in significant embedding distortion. On the other hand, in a two-dimensional hyperbolic space, it is possible to embed a tree with infinite depth while maintaining well-separated leaf nodes. Illustrated by the Poincar\'e disk model, which will be introduced in Section \ref{disk_model}, line segments in the right plot are of the same lengths as the hyperbolic distance metric grows exponentially near the edge of the disk. In fact, it has been studied in the literature that trees with infinity depth can be nearly isometrically embedded into two-dimensional hyperbolic spaces, whereas for Euclidean space the required dimension grows exponentially with the depth of tree \citep{krioukov2010hyperbolic}. Informally, this is because trees need exponential space for branching, and can be seen as a discrete hyperbolic space from a purely metric space perspective. Interestingly, many social networks or biological networks have local tree structures \citep{newman2018networks}.

Next, we introduce two popular parametric models for hyperbolic space that will be used throughout the paper, the Poincar\'e model and the hyperboloid model. While these models are equivalent in terms of isometry, they offer distinct advantages in terms of visualization and theoretical analysis, respectively. 

\subsubsection{Poincar\'e Disk Model}\label{disk_model}
The Poincar\'e disk model typically refers to the model for a two-dimensional hyperbolic space while it can be naturally extended to a higher-dimensional space. Points of a two-dimensional hyperbolic space with curvature $-K < 0$ are defined on the open unit disk of $\mathbb{R}^{2}$: $\mathbb{D}_2^{-K} = \{x \in \mathbb{R}^{2} ~|~ \|x\|_2 < 1\}.$ The corresponding distance metric of any two points $x,y \in \mathbb{D}_2^{-K}$, also known as the Poincar\'e metric, is defined as 
\begin{equation}
    d_{\mathbb{D}_2^{-K}}(x,y) = \frac{1}{\sqrt{K}} \Arc \left(1 + \frac{\|x - y\|_2^2}{(1 - \|x\|_2^2)(1 - \|y\|^2_2)}\right).
    \label{eq:poincare_metric_real}
\end{equation}
The Poincar\'e disk model can also be formulated in a simpler form on the complex plane:
\begin{equation}
    \mathbb{\widetilde D}_2^{-K} = \{z \in \mathbb{C} ~|~ |z| \leq 1\}, \quad d_{\mathbb{D}_2^{-K}}(z_1,z_2) = \frac{1}{\sqrt{K}}\operatorname{atanh}\left|\frac{z_1 - z_2}{1 - \bar{z}_1 z_2}\right|,
    \label{eq:poincare_metric_complex}
\end{equation}
where each point $(x, y) \in \mathbb{R}^2$ corresponds to $z = x + y i \in \mathbb{C}$ and $|z| = \sqrt{x^2 + y^2}$. As we shall see later, writing the model in the form of \eqref{eq:poincare_metric_complex} enables us to establish critical theoretical results for hyperbolic embedding by utilizing complex analysis techniques.

\begin{figure}[ht]
    \centering
    \includegraphics[width =0.5\linewidth]{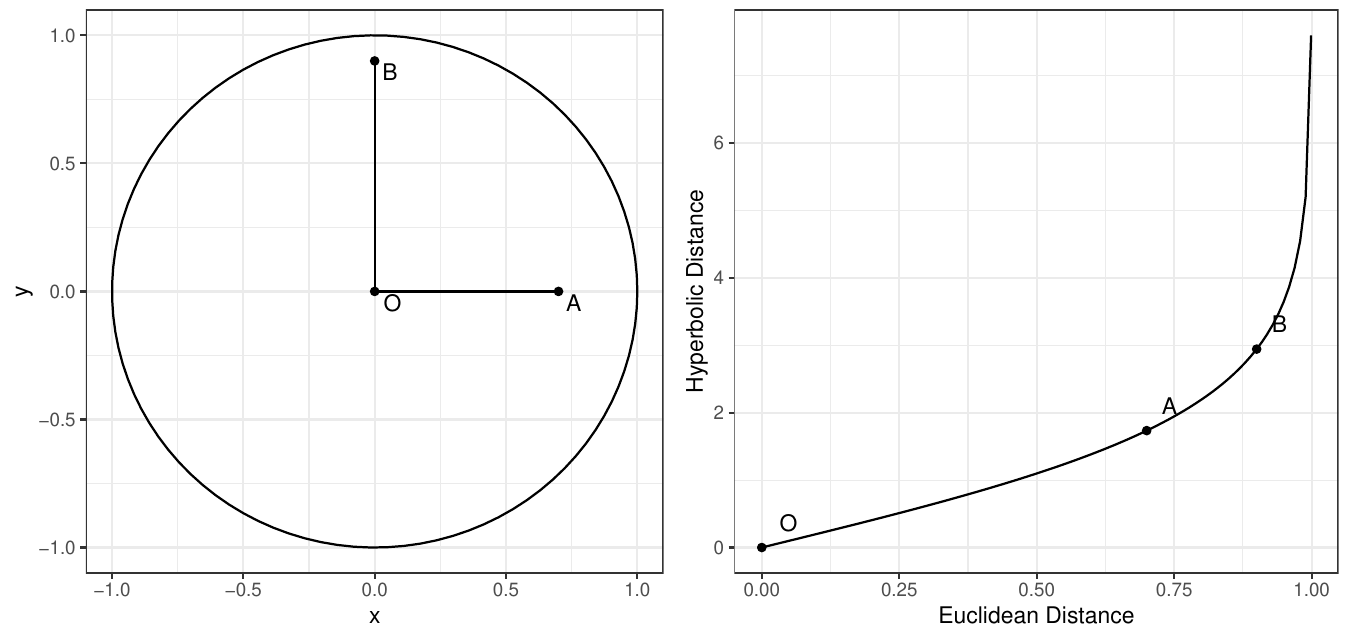}
    \caption{Left: Geodesics of $OA$ and $OB$ on the Poincar\'e disk model; Right: corresponding hyperbolic and Euclidean distances for OA and OB.}
    \label{fig:dist_to_Poincare_center}
\end{figure}

The Poincar\'e disk model is widely used for visualization purposes, because (1) all points lie within the open unit disk; and (2) the model is conformal, meaning angles in the Poincar\'e disk have identical values as if they were in the Euclidean space. However, when visualizing with the Poincar\'e disk model, one needs to keep in mind that the hyperbolic distance between a point and the center of the disk, grows exponentially compared with the Euclidean distance and the edge of the unit disk represents infinity. \Cref{fig:dist_to_Poincare_center} illustrates such property with a simple example of three points ($O$, $A$, and $B$) on the Poincar\'e Disk.

The model can be naturally extended to higher-dimensional hyperbolic spaces. For a $d$-dimensional hyperbolic space under Poincar\'e model, the set of points becomes open unit ball of $\mathbb{R}^d$ with the same form of distance metric as in \eqref{eq:poincare_metric_real}.


\subsubsection{Hyperboloid Model}
To introduce the hyperboloid model for a $d$-dimensional hyperbolic space, it would be convenient to first define the Lorentzian inner product. Define the $(d+1)\times(d+1)$ diagonal matrix $\Lambda = \text{diag}(1, 1, ..., 1, -1)$, the Lorentzian inner product is the vector product induced by $\Lambda$ on $\mathbb{R}^{d + 1}$, i.e., $x^T \Lambda y = \sum_{i = 1}^{d} x_i y_i - x_{d+1} y_{d + 1}$. Under the hyperboloid model, a $d$-dimensional hyperbolic space with curvature $-K<0$ is defined on the upper hyperboloid in $\mathbb{R}^{d+1}$, denoted as $\mathbb{H}_d^{-K} = \{x \in \mathbb{R}^{d+1}~| x^T \Lambda x  = -1, x_{d+1}>0\}$, with the corresponding distance metric to be
\begin{equation}
    d_{\mathbb{H}^{-K}_d}(x, y) = \frac{1}{\sqrt{K}}\Arc(-x^T \Lambda y).
    \label{eq:hyperboloid_metric}
\end{equation}
It should be clarified that although under the hyperboloid model, every point is represented by a $(d+1)$-dimensional coordinate, the effective dimension remains to be $d$ according to the constraint in the definition of $\mathbb{H}_d^{-K}$.

\begin{figure}
    \centering
    \includegraphics[width = 0.32\linewidth]{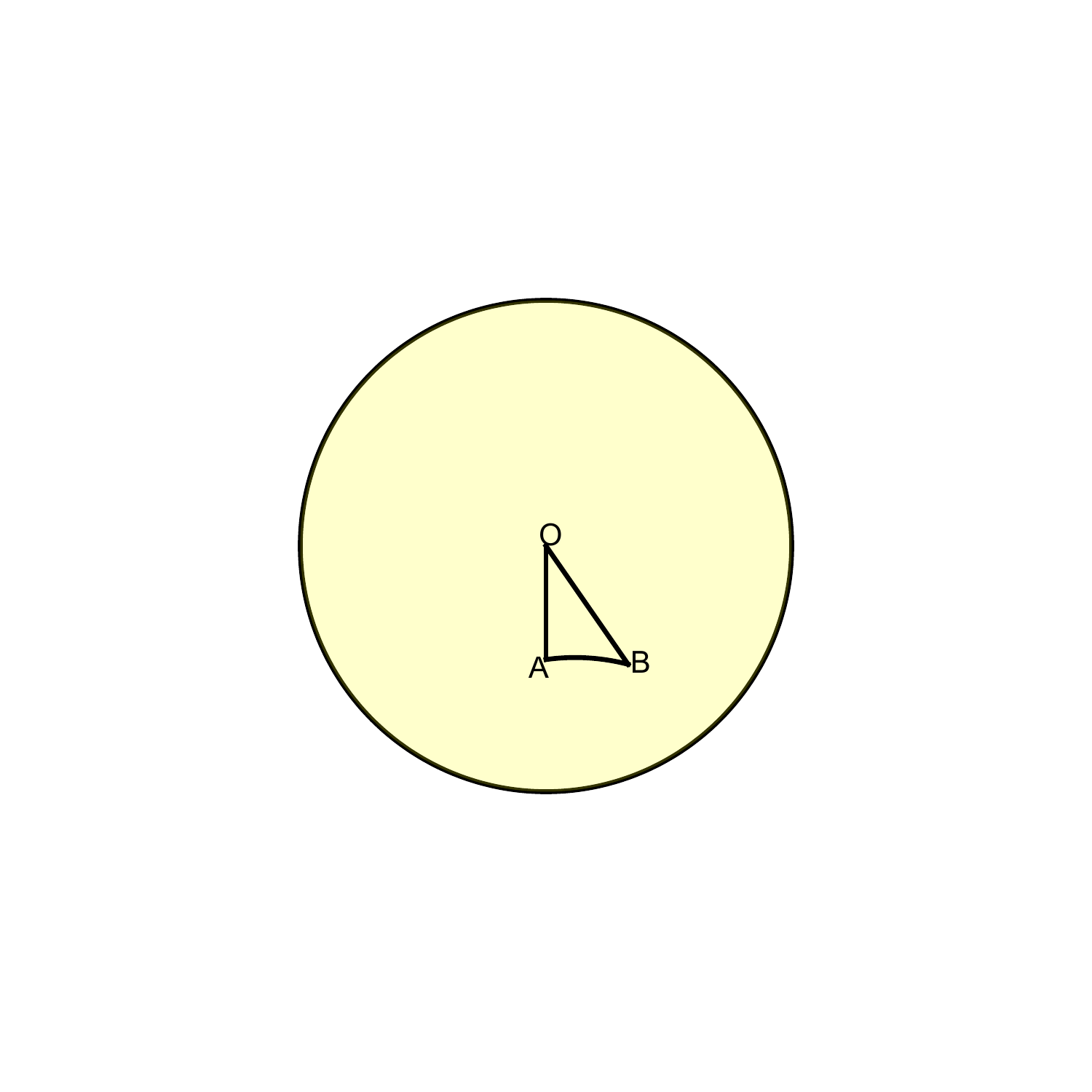}
    \includegraphics[width = 0.32\linewidth]{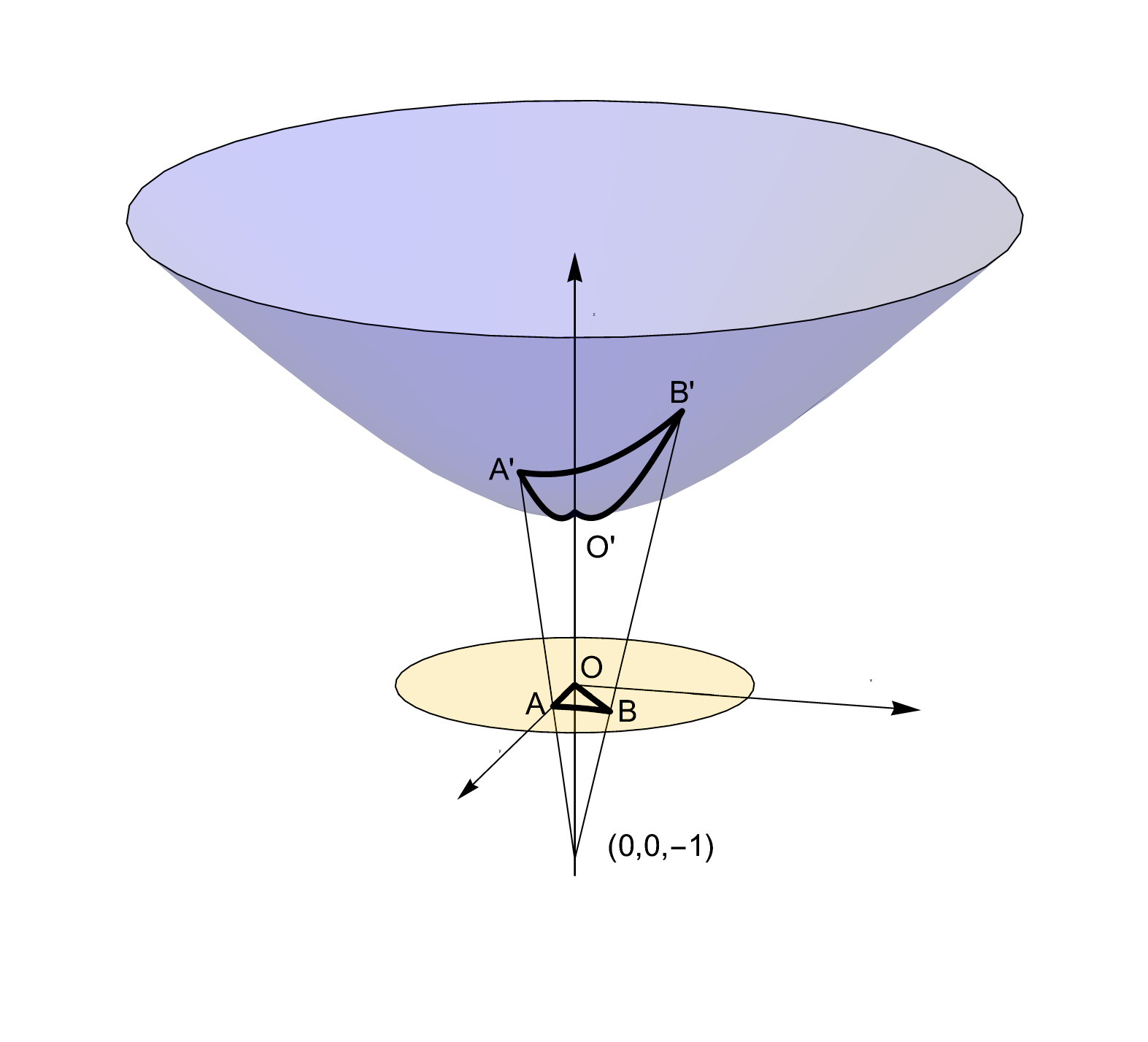}
    \includegraphics[width = 0.32\linewidth]{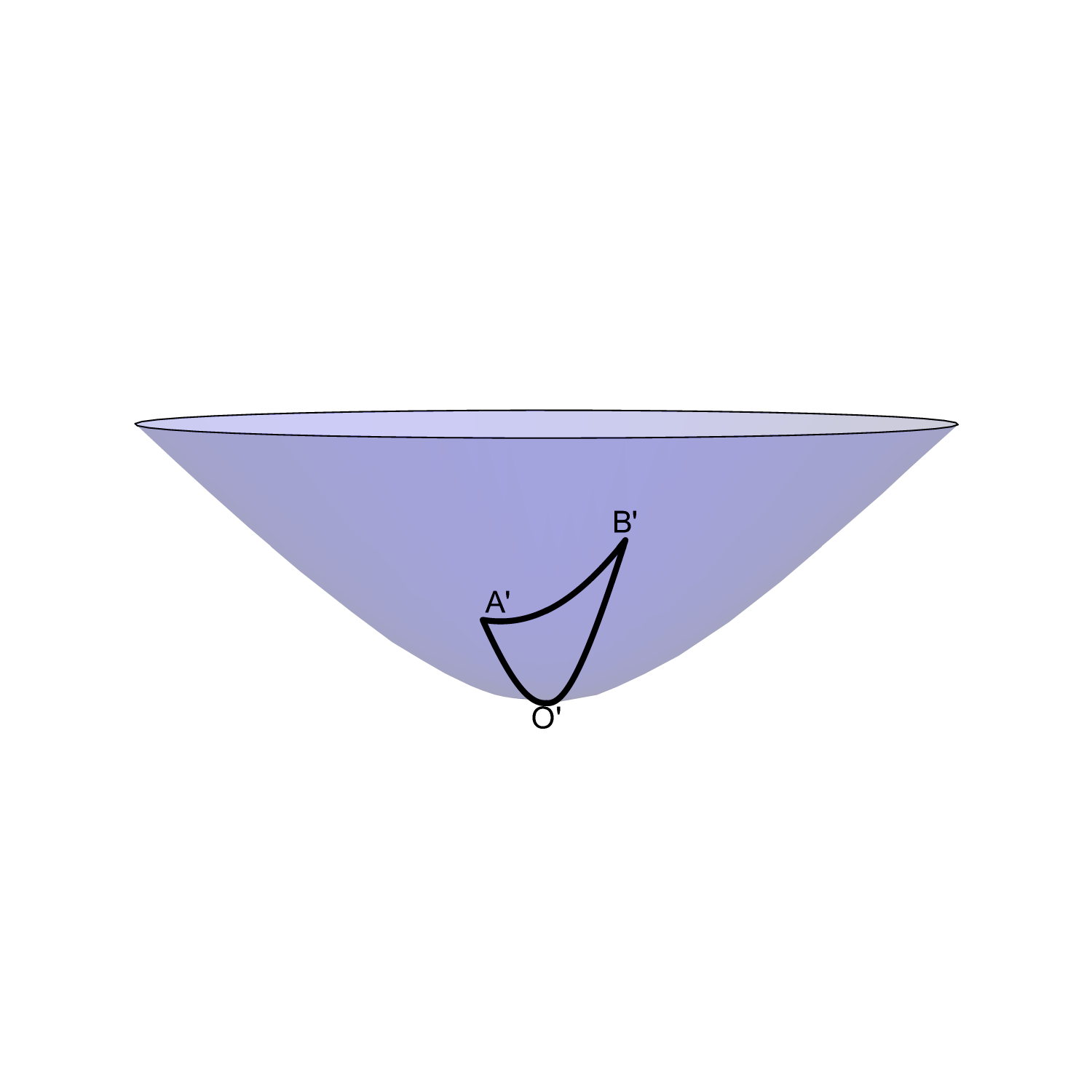}
    \caption{Left: a triangle on the Poincar\'e disk model; Middle: isometry illustration from hyperboloid model to Poincar\'e disk model; Right: a triangle on the hyperboloid model.}
    \label{fig:hyperboloid_poincare}
\end{figure}

\begin{remark}
The $d$-dimensional Poincar\'e ball model and hyperboloid model are equivalent, in the sense that there exists an isometry transformation $F$ from $\mathbb{H}_d^{-K}$ to $\mathbb{D}_d^{-K}$, $F: x^H \in \mathbb{H}_d^{-K} \to x^P \in \mathbb{D}_d^{-K}$. This transformation is given by $x_i^P = F(x_i^H) = x_i^H / (1 + x_{d+1}^H), 1 \leq i \leq d$. 
Conversely, the inverse transformation $F^{-1}$ is an isometry from $\mathbb{D}_d^{-K}$ to $\mathbb{H}_d^{-K}$. We illustrate such isometry with a $2$-dimensional example in \Cref{fig:hyperboloid_poincare}. The isometry $F$ can be interpreted as projecting a point from the hyperboloid (in blue) onto the unit disk (in yellow), with the projection source point $(0, 0, -1)$.
\end{remark}

\begin{remark}
Some literature \citep[e.g.,][]{chami2019hyperbolic} parameterize the hyperboloid model with the set of nodes and distance metric as $\mathbb{H}_d^{-K} = \{x \in \mathbb{R}^{d+1}\mid x^T \Lambda x = -1/K\}$ and $d_{\mathbb{H}^{-K}_d}(x, y) = \frac{1}{\sqrt{K}}\Arc(-K x^T \Lambda y)$. The two forms are equivalent as there exists the differentiable one-to-one mapping $\phi: x \to x / \sqrt{K}$ for all points between the two forms. Similar parametrization exists for the Poincar\'e disk model as well.
\end{remark}

Note that for both the hyperboloid model and the Poincar\'e disk model, hyperbolic spaces with different curvatures share a consistent definition set. The difference lies in the distance metrics, which have a scaling factor  $1 / \sqrt K$ that depends on the curvature parameters. Such property allows for a unified parametrization with arbitrary negative curvature parameters, which is crucial for model estimation and theoretical analysis. 

In the rest of the paper, we choose the hyperboloid model as the default model, which is favored over the Poincar\'e disk model for several technical reasons. From a theoretical analysis perspective, the pairwise distance matrix on the latent space can be fully characterized by products of low-rank matrices with non-linear transformations, which enables analysis techniques from the low-rank matrix completion literature to be applied. From a computational standpoint, optimizing a loss function that involves the distance metric under the hyperboloid model with a gradient-based method ends up with much simpler and stabler updating steps compared to the Poincar\'e disk model \citep{nickel2018learning}.

Below we present some important properties of hyperbolic spaces. For more technical details, we refer readers to \cite{c}. \Cref{property:impossibility_of_isometry} provides the fundamental argument for the identifiability and estimability of the curvature parameter. \Cref{property:isometry_of_hyperbolic_space}, similar to Euclidean self-isometries, characterizes the rotation and translation mappings in hyperbolic space.

\begin{property} 
Any two hyperbolic spaces with different curvatures, denoted as $\mathbb{H}_{d}^{-K}$ and $\mathbb{H}_{d'}^{-K'}$, with $K \neq K'$, are not isometric, irrespective of whether the dimensions $d=d'$ or $d\ne d'$. In other words, there does not exist any smooth one-to-one distance-preserved mapping $\phi: \mathbb{H}_{d}^{-K} \to \mathbb{H}_{d}^{-K'}$ such that for any $x, y \in \mathbb{H}_{d}^{-K}$, $d_{\mathbb{H}_{d'}^{-K}}(x, y) = d_{\mathbb{H}_{d'}^{-K'}}(\phi(x), \phi(y))$.
\label{property:impossibility_of_isometry}
\end{property}
\begin{property}
Every isometry of a $d$-dimensional hyperbolic space with curvature $-K<0$ to itself can be characterized with linear transformation $\mathcal{Q}$ under the hyperboloid model: $\mathcal{Q}: x \in \mathbb{H}_d^{-K} \to Qx \in \mathbb{H}_d^{-K}$, where $Q_{(d+1)\times(d+1)}$ is called a hyperboloid rotation matrix satisfying $Q\Lambda Q^T = \Lambda$. In the special case of $d = 2$, $Q$ must be a combination of the following three basic hyperbolic rotation matrices:
$$Q_1 = \left(\begin{array}{ccc}
1 & 0 & 0 \\
0 & \cosh \theta & \sinh \theta \\
0 & \sinh \theta & \cosh \theta
\end{array}\right),
~Q_2 = \left(\begin{array}{ccc}
\cosh \theta & 0 & \sinh \theta \\
0 & 1 & 0 \\
\sinh \theta & 0 & \cosh \theta
\end{array}\right),
~Q_3 = \left(\begin{array}{ccc}
\cos \theta & -\sin \theta & 0 \\
\sin \theta & \cos \theta & 0 \\
0 & 0 & 1
\end{array}\right).$$
\label{property:isometry_of_hyperbolic_space}
It should be clarified that although $Q$ is commonly referred to as hyperbolic rotation matrices, it actually includes both rotation and translation mappings in hyperbolic spaces.
\end{property}

\subsection{Model Setup}

We consider the network data where the edges are binary and undirected without any self-loop. Such networks, consisting of $n$ nodes, can be represented as a binary adjacency matrix $A_{n \times n}$, where $A_{ij} = 1$ if nodes $i$ and $j$ are connected or $A_{ij} = 0$ otherwise. Thus $A$ is symmetric and all diagonal entries are $0$. 
Following the latent space model framework, each node $i$ is assigned with a latent position $z_i$, where the latent space is a $d$-dimensional hyperbolic space with curvature $-K<0$. The latent positions are treated as fixed parameters that need to be estimated. Nevertheless, our estimation procedure and theoretical results remain valid for the situation where the latent positions are generated from some stochastic procedure, in the sense of estimating the realized values of the latent positions.

We propose a distance model with hyperbolic geometry for network data that has a unified form for hyperbolic spaces with different curvatures. Let the latent positions be $z_i \in \mathbb{H}_d^{-K}\subset \mathbb{R}^{d+1}$ for each $1 \leq i \leq n$.  Note that although $z_i$ can be viewed as a $(d+1)$-dimensional vector in $ \mathbb{R}^{d+1}$, it has an effective dimension of $d$. We use $Z_{n \times (d+1)}$ to denote the latent position matrix, where each row is the latent position $z_i$ for node $i$. The hyperbolic distance-based latent space model takes the form that for any $i < j$:
\begin{equation}
    A_{ij} \overset{ind.}{\sim} \mbox{Bernoulli}( P_{ij})~ \mbox{ with } {P_{ij}} = \sigma(d_{\mathbb{H}_d^{-K}}(z_i, z_j)),
    \label{eq:model}
\end{equation}
where $P_{ij}$ denotes the linkage probability between nodes $i$ and $j$, $d_{\mathbb{H}_d^{-K}}$ is given in \eqref{eq:hyperboloid_metric}, and $\sigma(x): [0, +\infty) \to [0, 1]$ is a smooth, strictly decreasing link function such that $0 \leq P_{ij} \leq 1$. Common choices of $\sigma(x)$ include logistic-type or exponential-type functions \citep{hoff2002latent,lubold2020identifying}. For convenience, we denote $P_{n \times n}$ as the linkage probability matrix and $\Theta_{n \times n}$ as the distance matrix whose diagonal entries are $0$ and off-diagonal entries are pairwise distances on the latent space, i.e., $\Theta_{ij} = \sigma^{-1}(P_{ij}) = d_{\mathbb{H}_d^{-K}}(z_i, z_j)$. Writing the proposed model in the matrix form, we have:
\begin{equation}
    \Theta = \frac{1}{\sqrt{K}}\Arc(-Z\Lambda Z^T) \mbox{ and }
    P = \sigma(\Theta),
    \label{eq:matrix_form_Theta}
\end{equation}
where $ \Arc(x)/{\sqrt{K}}$ and $\sigma(x)$ are applied to entries of $-Z\Lambda Z$ and $\Theta$, respectively. \Cref{eq:matrix_form_Theta} indicates that the pairwise distances are fully determined by the low-rank matrix $Z\Lambda Z^T$ under the hyperboloid model. Note that from \Cref{eq:matrix_form_Theta}, $K$ is directly related to network sparsity and various other empirical properties. It serves as a modeling parameter for network sparsity and heterogeneity, which will be further discussed in \Cref{sec:simulation_network_properties}.

\begin{figure}[ht]
    \centering
    \includegraphics[width=\linewidth]{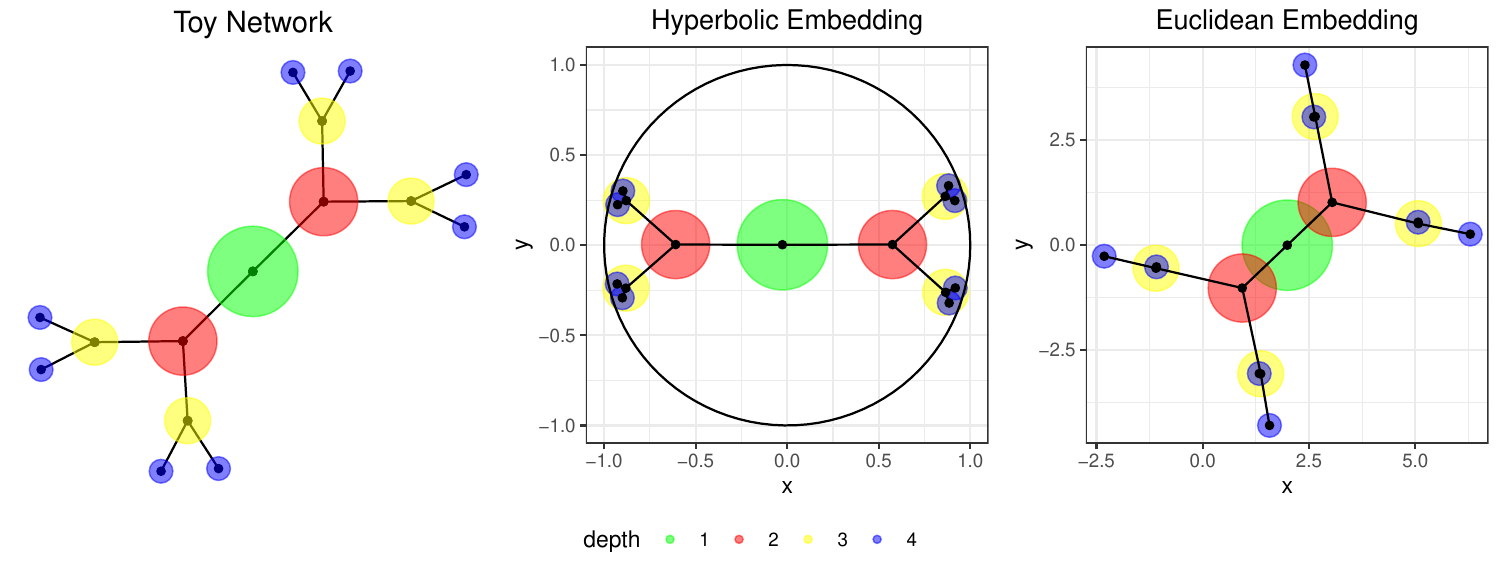}
    \includegraphics[width=1.025\linewidth]{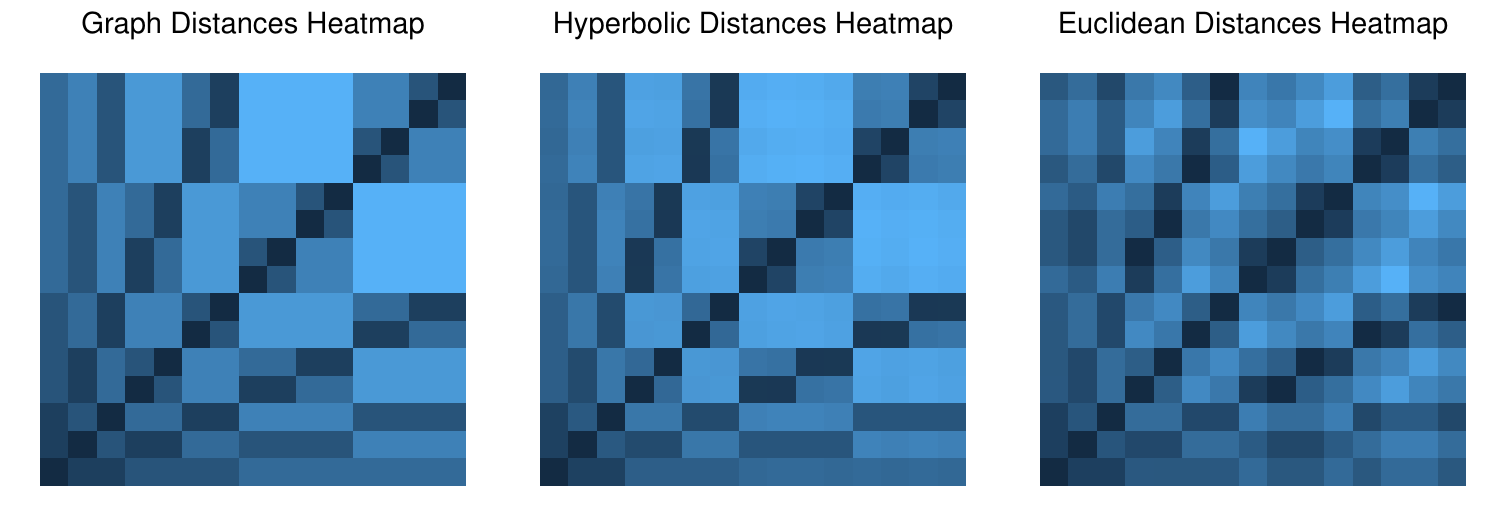}
    \caption{A toy network with visualization of latent positions and distance heatmap.}
    \label{fig:toy_example}
\end{figure}

We demonstrate the advantage of the proposed hyperbolic model over the Euclidean model using a toy example in \Cref{fig:toy_example}. In this example, the network is a $2$-ary tree with a depth of 4. We fit two latent space models with hyperbolic and Euclidean distance metrics, respectively, with the link function being $\sigma(x) = 2\text{logistic}(-x) = 2/(1 + e^{x})$. 
As shown in the figure, the hyperbolic embeddings, i.e., the latent positions learned from the hyperbolic model, accurately capture the hierarchical structure of the original network. In contrast, the Euclidean embeddings result in overlapping positions for four blue nodes with the yellow nodes. We further illustrate the heat maps of distance matrices learned from the two models and compare them to the shortest path distance of the original network. It is observed that hyperbolic embeddings recover more consistent pairwise distances than Euclidean embeddings. In the Supplementary Materials (Section B.3), we further provide examples demonstrating the advantage of hyperbolic models for embedding networks with community structures.

\subsection{Identifiability}
\label{sec:identifiability}

To ensure the consistency of parameter estimation, it is crucial to investigate whether the curvature parameter $K$ and latent position matrix $Z$ can be uniquely determined given the probability matrix $P$. By choosing a strictly decreasing link function $\sigma(x)$, we can directly focus on establishing identifiability conditions for $\Theta$. In this context, we assume that the dimension of the latent space is fixed, leaving the discussion of related selection problems for future consideration.

If the curvature parameter $K$ is fixed, the geometry of latent space is fully known. Combining \Cref{property:isometry_of_hyperbolic_space} with \Cref{eq:matrix_form_Theta}, we observe that $Z$ can only be identified up to an arbitrary hyperbolic rotation matrix $Q_{(d+1)\times (d+1)}$. Specifically, from $Q\Lambda Q^T = \Lambda$ we have $(ZQ) \Lambda (ZQ)^T = Z \Lambda Z^T$. Since equation $Q \Lambda Q^T = \Lambda$ has $(d+1)(d+2)/2$ effective constraints on entries of $Q$, we generally need $(d+1)d/2$ additional constraints to remove the indeterminacy of $Z$. In practice, a convenient set of identifiability constraints is to enforce $Z^TZ$ to be diagonal. Let $U S U^T$ represent the eigen-decomposition of $Z \Lambda Z^T$, we know that $S$ contains exactly $d$ positive eigenvalues and one negative eigenvalue \citep{tabaghi2020hyperbolic}. Thus, we can set $Z = U |S|^{1/2} \Lambda$, where the diagonal entries of $|S|^{1/2}$ are the square root of corresponding absolute values of eigenvalues in $S$.

In terms of identifying the curvature parameter $K$, \Cref{property:impossibility_of_isometry} indicates that smooth distance-preserved mapping from one hyperbolic space to the other does not exist. However, this statement needs closer examination when considering a set of $n$ discrete points in the latent space. The problem of identifying $K$ is equivalent to show the \textit{non-existence} of the isometric mapping $\phi: \{z_i\}_{i=1}^{n} \subset \mathbb{H}_{d}^{-K} \to \{z'_i\}_{i=1}^{n} \subset \mathbb{H}_{d}^{-K'}$, such that $d_{\mathbb{H}_{2}^{-K}}(z_i, z_j) = d_{\mathbb{H}_{2}^{-K'}}(\phi(z_i), \phi(z_j))$ holds for all $i \neq j$ when $K \neq K'$. Combined with \Cref{eq:hyperboloid_metric}, we can rewrite the set of equations as 
$    \Arc(-z_i^T\Lambda z_j) = d_{\mathbb{H}_{d}^{-1}}(z_i, z_j) = \sqrt{K/K'} d_{\mathbb{H}_{d}^{-1}}(\phi(z_i), \phi(z_j)) = \sqrt{K/K'} \Arc(-{z_i'}^T\Lambda {z_j'}).
$
It is challenging to rigorously characterize the solution because of the non-linearity of equations and the singularity issue of discrete geometry. In irregular cases, such as when all latent positions are identical or lie on a one-dimensional line, the curvature is not well-defined as the latent space becomes degenerated. However, intuitively when $n$ is sufficiently large we would not encounter singularity issues. Thus, the focus of our discussion is on regular cases where discrete points are appropriately distributed in the latent space. 
 In general, solutions exist only when the number of unknowns, $n(d + 1)$, is greater than the total number of equations and constraints, $n(n + 1)/2 + (d+1)d/2$, where we include the identifiability constraints on $Z$ as well as the constraints of hyperboloid model. Solving this quadratic relationship, we can conclude that the equations above have solutions only when $n \leq d + 1$, which indicates: (1) in general, $d + 2$ discrete points would be sufficient to determine the curvature of a hyperbolic space; and (2) when $K \neq K'$, isometric mappings are possible only when the subspace which discrete points lie on degenerates.
To further illustrate this, we provide an illustrative example to demonstrate that 4 nodes are sufficient to distinguish the curvatures of different 2-dimensional hyperbolic spaces in the Supplementary Materials. 

\section{Estimation Procedure}
\label{sec:estimation_and_initialization}
We   estimate the model parameters by minimizing the negative log-likelihood function:
\begin{equation*}
    \mathcal{L}(K, Z) = -  \sum_{i \neq j}[A_{ij} \log(\sigma\left(\Theta_{ij})\right) + (1-A_{ij})\log\left(1 - \sigma(\Theta_{ij})\right)], ~\mbox{s.t.}~ z_i^T \Lambda z_i = -1, 1 \leq i \leq n,
\label{eq:loss}
\end{equation*}
where $\Theta_{ij} = \Arc(-z_i^T \Lambda z_j) / \sqrt{K}$. This optimization problem associated with both $K$ and $Z$ is challenging due to the non-convex function form and the constraints on $Z$. To address these challenges, we develop a gradient-based optimization algorithm outlined in \Cref{alg:mgd_gd}. In each iteration, the algorithm updates $K$ with Gradient Descent (GD) and updates $Z$ with Manifold Gradient Descent (MGD) simultaneously.  We propose using different step sizes for updating $K$ and $Z$ to ensure stability, with each step size normalized by the effective sample sizes when calculating the corresponding gradients. In particular, we choose $\eta_K = O(K^0/ n^2)$ and $\eta_Z = O(1 / n)$, respectively. Notably, our numerical results show that this choice of $\eta_K$ also prevents the updated $K$ from being negative and invalid in \Cref{alg:mgd_gd}.

\begin{algorithm}
\caption{The Combined Gradient-based Algorithm} \label{alg:mgd_gd}
\begin{algorithmic}[1]
\Require {Adjacency matrix: $A$; dimension: $d$; initial values: $K^0, Z^0$; step sizes: $\eta_K, \eta_Z$; threshold: $\epsilon$.}

\State Set $\delta = 2 \epsilon$, $t = 0$;
\While{$\delta > \epsilon$}
\State $K^{t + 1} = K^{t} - \eta_K \times \nabla_{K} \mathcal{L}(K^{t}, Z^{t}; A, d)$;
\State $Z^{t + 1} = \text{MGD}(A, d, \eta_Z, K^{t}, Z^{t}$); \Comment{See \Cref{alg:mgd_alone}.}
\State Set $\delta = \mathcal{L}(K^{t}, Z^{t}) - \mathcal{L}(K^{t+1}, Z^{t+1})$, $t = t + 1$;
\EndWhile

\Ensure {$\hK = K^t$, $\hZ = Z^t$.}
\end{algorithmic}
\end{algorithm}

Utilizing Riemannian optimization techniques \citep{nickel2018learning}, MGD directly updates latent positions along the gradient direction on the hyperboloid manifold. Outlined in \Cref{alg:mgd_alone}, MGD first derives the direction of the steepest descent from the Euclidean gradient and then calculates the Riemannian gradient on the tangent space $\mathcal{T}_z = \{x\in \mathbb{R}^{d+1} | x^T \Lambda z = 0\}$ with projection mapping \eqref{eq:proj_mapping}. In the second step, MGD updates the latent position along the geodesic on $\mathbb{H}_d^{-K}$ with the Riemannian gradient direction, using the exponential map \eqref{eq:exp_mapping}. Note that steps 4 and 5 inside the for-loop can be updated with parallel computing to accelerate computation. 
\begin{align}
    \text{proj}_{z}(g)&: \mathbb{R}^{d+1} \to \mathcal{T}_z \quad \mbox{s.t.} \quad \text{proj}_{z}(g) = g + (g^T \Lambda z) \cdot z, \label{eq:proj_mapping}\\
    \text{exp}_{z}(g)&: \mathcal{T}_z \to \mathbb{H}_d^{-K} \quad \mbox{s.t.} \quad  \text{exp}_{z}(g) = \cosh(\sqrt{g^T\Lambda g})\cdot z + \sinh(\sqrt{g^T \Lambda g})/\sqrt{g^T \Lambda g} \cdot g. \label{eq:exp_mapping}
\end{align}
Comparing optimizing $Z$ with ad-hoc gradient descent algorithms, MGD is able to utilize the natural geometry of the manifold, leading to faster convergence speed and better numerical stability. Finally, the output of \Cref{alg:mgd_gd} can be further transformed to meet the identifiability conditions described in \Cref{sec:identifiability}. This can be achieved by performing eigen-decomposition $Z \Lambda {Z}^T = U S U^T$ and set $\hat Z = U |S|^{1/2} \Lambda$.

\begin{algorithm}
\caption{The MGD Update Algorithm} \label{alg:mgd_alone}
\begin{algorithmic}[1]
\Require {Adjacency matrix: $A$; dimension: $d$; step size: $\eta_{Z}$; estimators: $K^t, Z^t$.}
\State Set $\Lambda_{(d+1) \times (d+1)} = diag(1, 1, ..., 1, -1)$;
\State $g^{E} = \Lambda \nabla_{Z}\mathcal{L}(K^t, L^t; A, d)$; \Comment{Calculate the steepest direction.}
\For{$i=1, 2, ...n$}
\State $g^{R}_i = \text{proj}_{z^t_i}(g^{E}_i)$; \Comment{Calculate the Riemannian gradient.}
\State $z_i^{t+1} = \exp_{z_i^t}(-\eta_Z ~ g^{R}_i)$; \Comment{Update along the geodesic.}

\EndFor
\Ensure {$Z^{t+1} = \left(z_i^{t+1}\right)_{i=1}^n$.}
\end{algorithmic}
\end{algorithm}

Since the optimization problem is non-convex, initialization plays an important role in finding the global minimum and the corresponding optimal estimators. We use $P^0$, $Z^0$, and so on, to denote the initial values of these parameters. Following \cite{ma2020universal}, we develop two initialization methods based on the Universal Singular Value Thresholding (USVT, \citealp{chatterjee2015matrix}). Intuitively, USVT produces a low-rank estimation $P^0$ of the probability matrix by taking SVD on $A$. By applying the inverse of link function $\sigma(x)$ on $P^0$, we are then able to derive an initial estimation $\Theta^0$ of the distance matrix. Nevertheless, solving both $K^0$ and $Z^0$ from $\Theta^0$ remains to be a challenging problem due to the non-convexity.
If $K$ is known, according to \Cref{eq:matrix_form_Theta} an initial estimation of $G_0 = Z^0 \Lambda {Z^0}^T$ can be obtained with an entry-wise inverse mapping. Then $Z^0$ can be obtained with the eigen-decomposition of $G_0$. Denote $G_0 = USU^T$ to be the eigen-decomposition of $G_0$, we define $\tilde{Z}_0 = \tilde{U} \tilde{|S|}^{1/2} \Lambda$ to be a low-rank approximation, where $\tilde{S}$ contains the $d$ leading positive eigenvalues and the smallest negative eigenvalue and $\tilde{U}$ contains the corresponding eigenvectors. Finally, the initial estimation $Z^0$ can be obtained by projecting $\tilde{Z}$ onto $\mathbb{H}_d^{-K}$ using an established formula in the literature \citep{tabaghi2020hyperbolic}. In practice, when we do not have access to $K$, we propose to first fit the model with several initial candidate values of $K^0$ while fixing $\eta_K = 0$, and choose the value with the lowest loss as $K^0$ and run \Cref{alg:mgd_gd}. The initialization method is summarized in \Cref{alg:init_w_USVT}. 

\begin{algorithm}
\caption{Initialization with USVT} \label{alg:init_w_USVT}
\begin{algorithmic}[1]
\Require{Adjacency matrix: $A$; dimension: $d$; threshold: $\tau$; candidates list of $K^0$: $\{K_1, K_2,...,K_m\}$.}
\State Set $\Lambda_{(d+1) \times (d+1)} = diag(1, 1, ..., 1, -1)$;
\State $P^0 = \sum_{\sigma_i \geq \tau}\sigma_i u_i v_i^T$, where $\sum_{1 \leq i \leq n}\sigma_i u_i v_i^T$ is SVD of $A$; \Comment{USVT} 
\State $\Theta^0 = \sigma^{-1}[(P^0 + {P^0}^T) / 2]$;
\For{$j = 1,2,..., m$}
\State $G_0 = \cosh(\sqrt{K_j} \Theta^0)$, $\tilde{Z}_j = \text{Project}(\tilde{U} \tilde{|S|}^{1/2} \Lambda$);
\State ${Z}_j = \text{Algorithm 1}(A, d, K_j, \tilde{Z}_j, 0, \eta_Z, \epsilon)$;
\EndFor
\State $j_0 = \text{argmin}_{1\leq j \leq m} \mathcal{L}(K_j, \hat{Z}_j)$
\Ensure $K^0 = K_{j_0}$, $Z^0 = \tilde{Z}_{j_0}$.
\end{algorithmic}
\end{algorithm}

In simulation studies, we find that when the latent space has large $K$ (highly curved), USVT does not always produce a good initial estimation $\Theta^0$, leading to an unsatisfactory estimation $Z^0$ (see \Cref{sec:simulation_compare_initialization}). To address this challenge, we propose a refined initialization procedure: after step 3 and before step 4 in \Cref{alg:init_w_USVT}, we fit a higher-dimensional Euclidean distance model to obtain a better estimation $\Theta^0$, which will be used in steps 4-8.

\begin{remark}
Although higher-dimensional Euclidean models can be useful for the purpose of initialization, it is important to note that they are not necessarily superior to lower-dimensional hyperbolic models. This will be illustrated in the simulation studies,
where we demonstrate that a higher-dimensional Euclidean model can easily overfit when the curvature of the latent space is close to zero. Conversely, when the curvature parameter $K$ is large, even higher-dimensional Euclidean models are unable to accurately estimate the probability matrix.
\end{remark}



\section{Theoretical Analysis}

In this section, we first utilize hyperbolic geometry techniques to characterize the error bound for the embedded distance matrix of latent positions from hyperbolic spaces with different curvatures, which reveals the importance of the latent space curvature for general distance-based hyperbolic embedding methods. 
We then present the consistency rates for the maximum likelihood estimators of both the curvature and the latent positions.

We use $P$, $Z$, and similar symbols, to denote the true values of parameters, and use $\hat P$, $\hat Z$, and so on, to represent their estimators. The following assumptions are made throughout.

\renewcommand{\theenumi}{\Roman{enumi}}
\begin{enumerate}
\label{item:assumptions}
    \item Values of both the estimators and true model parameters belong to the bounded parameter space 
$
       \Xi = \left\{ 0< C_1^{-1} \leq K \leq C_1, 0 < C_2 \leq |\Theta_{ij}|  \leq C_3 < \infty \right\},
$
    where $C_1$, $C_2$, and $C_3$ are positive constants.
    \item The link function satisfies $\sigma(x): [0, +\infty) \to [0, 1]$ and $\sigma'(x) <0$ is continuous for $C_2 \leq x \leq C_3$.
    \item The limiting matrix $\lim_{n \to \infty} Z^TZ/ n = \Sigma_Z$ exists. Furthermore, $\Sigma_Z$ is diagonal and contains distinct non-zero eigenvalues.   
\end{enumerate}

In Assumption I, we assume the model parameters and the estimators belong to a compact set. The lower bound on the pairwise distance of latent positions is necessary to avoid the irregular cases such as all latent positions are identical, which has been discussed in the model identifiability. Later when presenting theoretical results, we will provide more discussions on how the smallest distance assumption could be relaxed by alternative weaker assumptions. The assumption on the upper bound of the pairwise distance and its implication to network sparsity will be discussed with more details in \Cref{sec:discussion}. Assumption II imposes regularity and smoothness conditions on the link function, which is satisfied by many popular choices such as logistic-type functions. Assumption III ensures the identifiability of $Z$, which could be replaced by other $(d+1)d/2$ constraints on $Z$. However, the stated assumption is convenient to work with in practice. 

\subsection{Embedding Error Caused by Misspecified Curvature}
\label{sec:embedding_error}

Motivated by \Cref{property:impossibility_of_isometry}, which states that misspecifying curvature leads to the non-existence of distance-preserved mapping, we are interested in characterizing the error bound of the distance matrix for $n$ discrete nodes. Here we focus on the two-dimensional case and formulate the problem as follows. Given (1) two hyperbolic spaces $\mathbb{H}_2^{-K}$ and $\mathbb{H}_2^{-K'}$ with $K \neq K'$, and (2) a set of discrete points $\{z_i\}_{i = 1}^n \subset \mathbb{H}_2^{-K}$, what is the lower error bound of the following approximation in \Cref{eq:embedding_problem}? 
\begin{equation}
\min_{\{z_i'\}_{i = 1}^n \subset \mathbb{H}_2^{-K'}} ||\Theta - \Theta'||_F^2 = \min_{\{z_i'\}_{i = 1}^n \subset \mathbb{H}_2^{-K'}} \sum_{i \neq j}\left[d_{\mathbb{H}_2^{-K}}({z_i, z_j}) - d_{\mathbb{H}_2^{-K'}}(z_i', z_i')\right]^2
\label{eq:embedding_problem}
\end{equation}

Note that \Cref{eq:embedding_problem} considers the error of pairwise distance matrix, whose value is invariant to the choice of parameterization of hyperbolic geometry. Thus, we can replace the distance metric in \Cref{eq:embedding_problem} to be the metric defined in \Cref{eq:poincare_metric_complex} under the Poincar\'e disk model on the complex plane, $\mathbb{\widetilde D}_2$. Utilizing the Schwarz-Pick Theorem \citep{osserman1999schwarz} in complex analysis, we can establish the following lower bound:

\begin{theorem}\label{thm:embedding_error}
Consider (1) two hyperbolic spaces $\mathbb{H}_2^{-K}$ and $\mathbb{H}_2^{-K'}$ with $K \neq K'$, and (2) a set of discrete points $\{z_i\}_{i = 1}^n \subset \mathbb{H}_2^{-K}$. Suppose Assumption I holds and additionally, the following smoothness assumption holds: (1) when $K<K'$, there exists a holomorphic mapping $\phi: \mathbb{\widetilde D}_2 \to \mathbb{\widetilde D}_2$ s.t. $\phi(z_i) = z_i'$; or (2) when $K' < K$, there exists a holomorphic mapping $\psi: \mathbb{\widetilde D}_2 \to \mathbb{\widetilde D}_2$ s.t. $\psi(z_i') = z_i$. Then we have:
\begin{equation*}
    \frac{1}{n(n-1)}\|\Theta - \Theta'\|_F^2 \geq \frac{C_2^2}{C_1} \left(1 - \min \left\{\sqrt{\frac{K'}{K}}, \sqrt{\frac{K}{K'}}\right\}\right)^2.
\end{equation*}
\end{theorem}
Note that a holomorphic function in complex analysis is defined as a complex differentiable mapping such that it can be locally approximated by Taylor expansion. Such smoothness assumption aligns with the nature of the minimization problem in \Cref{eq:embedding_problem}, where the loss function itself is inherently smooth. In the context of model optimization where $\Theta$ denotes the true parameters and $\Theta'$ denotes the estimators, the holomorphic assumption essentially means that the likelihood function is changing smoothly within a small neighborhood of the true latent positions. 

\Cref{thm:embedding_error} implies that for any distance-based hyperbolic embedding method, the average embedding error of the pairwise distance is at least of order $1 - \min \{\sqrt{{K'}/{K}}, \sqrt{{K}/{K'}}\} $, which indicates that 
there exists $M>0$ such that $\frac{1}{n(n-1)}\|\Theta - \Theta'\|_F^2 \geq M |K - K'|$. 

An important implication of Theorem \ref{thm:embedding_error} is as follows. Instead of setting $K=1$ as most existing hyperbolic embedding methods do in practice, to improve the embedding quality we ought to estimate the curvature of latent space. Moreover, \Cref{thm:embedding_error} indicates that the latent space curvature has a unique, irreplaceable effect on the characteristics of the distance matrix of latent positions, as the conclusion holds without any distributional assumption over the generating mechanism of latent positions. 

\begin{remark}
In \cite{chami2019hyperbolic}, the authors proposed the Hyperbolic Graph Convolutional Network (HGCN) model that learns hyperbolic embeddings, finding that HGCN achieved much better performance on downstream tasks with learnable curvature parameters other than fixing them as $-1$. While the authors illustrated such improvement from a numerical perspective, our \Cref{thm:embedding_error} provides a theoretical justification. 
Furthermore, while Theorem 4.1 in \cite{chami2019hyperbolic} states that for distance-based linear classifiers there always exist equivalent sets of embeddings from hyperbolic spaces with different curvatures, our \Cref{thm:embedding_error} further indicates that the best embeddings can only be obtained when the curvature is optimal. 
\end{remark}

\begin{remark}
It can be proved that the minimum distance assumption in Assumption I, i.e., $\min_{i \neq j}|\Theta_{ij}| \geq C_2$, can be replaced by a weaker assumption: $\|\Theta\|_F / \sqrt{n} \geq C_2$, which imposes a lower bound on the average pairwise distances. The conclusion still holds because the Frobenius norm error bound in \Cref{thm:embedding_error} tries to characterize the average error instead of the extreme values. Thus, even if a few latent positions are overlapped, the average error rate would not change as long as the overall pairwise distances are well-behaved. 
\end{remark}

\subsection{Consistency Results}
In this section, we present the consistency results of the maximum-likelihood estimators. Developing consistency results for the maximum-likelihood estimators with hyperbolic geometry is a challenging problem due to the non-linearity and non-convexity of the hyperbolic distance metric. 
We establish consistency results for two situations: (1) the consistency rates of $\hat P$ and $\hat \Theta$, and 
(2) the consistency rates of $\hK$ and $\hZ$, where the latent space dimension is correctly specified. 

We first characterize the consistency of $\hat P$ and $\hat \Theta$ 
following the approximate low-rank framework in the matrix completion literature \citep{davenport20141}. In our case, we refer to the matrices as having a bounded nuclear norm: $\|Z\Lambda Z^T\|_* \leq n \sqrt{(d+1)} \cosh(\sqrt{C_1} C_3)$. Here, $\|\cdot\|_*$ represents the summation of the singular values of a matrix. 

\begin{theorem}[Consistency of $\hat \Theta$ and $\hat P$]
\label{thm:consistency_theta}
Suppose Assumptions I-III hold, 
then we have with probability at least $1 - \tilde{C}_1 / n$, 
\begin{align*}
    \frac{1}{n(n-1)}\|\hP - P\|_F^2 &\leq \tilde{C} \cosh(\sqrt{C}_1 C_3) \alpha_\sigma \sqrt{\frac{d+1}{n}}, \\
    \frac{1}{n(n-1)}\|\hT - \Theta\|_F^2 &\leq \frac{\tilde{C} \cosh(\sqrt{C_1}C_3) \alpha_\sigma}{\min_{C_2 \leq x \leq C_3} |\sigma^{\prime}(x)|^2}  \sqrt{\frac{d+1}{n}},
\end{align*}
where $d$ is the true latent space dimension, $\alpha_{\sigma} = \sup_{C_2\leq x \leq C_3} \frac{|\sigma^{\prime}(x)|}{\sigma(x) (1 - \sigma(x))}$, $\tilde{C}$ and $\tilde{C}_1$ are absolute constants that are independent of any other parameters.
\end{theorem}

\Cref{thm:consistency_theta} indicates that both $\|\hP - P\|_F^2 / [n(n-1)]$ and $\|\hT - \Theta\|_F^2 /[n(n-1)]$ have consistency rates of $O_p(1/\sqrt{n})$. Note that the approximate low-rank condition can be induced from Assumption I, since $\|\hZ \Lambda \hZ^T\|_* \leq \sqrt{(d+1)}\|\hZ \Lambda \hZ^T\|_F \leq  n \sqrt{(d+1)} \cosh(\sqrt{C_1} C_3)$. 



\Cref{thm:consistency_theta} holds for any latent space dimension $d \geq 2$. Combining \Cref{thm:embedding_error} and \Cref{thm:consistency_theta}, we are then able to characterize the consistency rates of $\hK$ and $\hZ$ when the dimension of the latent space is $d=2$. To do so, we would first characterize the consistency rate of $\hZ \Lambda \hZ^T$ and then derive the rate for $\hZ$ with additional regularity assumptions on the singular values of $Z$. 

\begin{theorem}[Consistency of $\hK$ and $\hZ$]
\label{thm:consistency_KZ}
Given $d = 2$, suppose Assumptions I-III and the assumptions stated in \Cref{thm:embedding_error} hold, then we have
$
    |\hK - K|^2 = O_p(1/\sqrt{n})$ and $
    \|\hZ - Z\|_F^2 / n = O_p(1/\sqrt{n}).
$
\end{theorem}

\begin{remark}
     In \cite{shalizi2017consistency}, the authors considered a related problem regarding the consistency results of maximum-likelihood based estimators while making a strong asymptotic assumption on the link probability, which does not fit with our model setup. Furthermore, our results have the following advantages: (1) in \cite{shalizi2017consistency}, the consistency results of latent positions are derived assuming the latent space curvature is known, while we consider a more practical situation where both the latent positions $Z$ and the curvature parameter $K$ need to be estimated; and (2) the analytic approach used in \cite{shalizi2017consistency} is not able to characterize the consistency rates of estimators, while in \Cref{thm:consistency_theta} and \Cref{thm:consistency_KZ} we derive the consistency rates with techniques in low-rank matrix completion literature.
\end{remark}

\begin{remark}
The consistency rates derived in \Cref{thm:consistency_KZ} can be naturally generalized to higher dimensions if \Cref{thm:embedding_error} holds for any $d \geq 2$. Examining the statement of \Cref{thm:embedding_error}  for higher-dimensional hyperbolic spaces is itself an interesting but challenging mathematical problem in hyperbolic geometry and is left for further exploration.
\end{remark}


\section{Simulation Studies}
\label{sec:simulation}

In this section, we use simulation studies to assess the properties and performance of the proposed model. We focus on the following aspects: (1) the effect of latent space curvature on the properties of generated networks; (2) the estimation consistency of the maximum-likelihood estimators; (3) the impact of latent space curvature on model fitting; and (4) power analysis and coverage rate analysis of a proposed inference procedure on the curvature parameter.
We employ a general setup where networks are generated based on the model defined in \eqref{eq:model} with a $2$-dimensional hyperbolic space. The link function is a logistic-type function $\sigma(x) = 2 \text{logistic}(-x) = 2/(1 + e^{x})$, such that the edge connecting probability is $1$ when the distance is $0$. Latent positions are generated from $i.i.d.$ samples following uniform distribution on the hyperbolic disk with radius $3$, such that the edge connecting probability has a lower bound of around $0.005$. More details of the uniform distribution on the hyperbolic disk as well as the derivation of the sampling procedure can be found in the Supplementary Materials.

\subsection{Model Properties versus Latent Space Curvature}
\label{sec:simulation_network_properties}

We first illustrate the geometric effect of the latent space curvature on the generated networks. From both experiments presented in the Supplementary Materials (Section B.1), we can see that the curvature parameter has a great effect on the global and local properties of generated networks. Specifically, a larger $K$ leads to lower edge density and more heterogeneous network structures. Therefore, in practice, $K$ can be interpreted as a modeling parameter for sparsity and heterogeneity, where larger values of $K$ indicate sparser and more heterogeneous networks.

\subsection{Consistency of the Maximum-Likelihood Estimators}
\label{sec:simulation_compare_initialization}

In this experiment, we examine the consistency of maximum-likelihood estimators. We vary the curvature parameter $K \in \{0.5, 1, 1.5\}$ and sample size $n \in \{250, 500, 1000, 2000\}$. The candidate initial values of $K^0$ are $\{0.1, 1, 10\}$. The step sizes for \Cref{alg:mgd_gd} are $\eta_K = 1 / n^2$ and $\eta_Z = 1/n$, which are scaled according to the effective sample size. We use relative estimation error as evaluation criteria, defined as $\Delta K = |\hK - K|$, $\Delta Z = \|\hZ - Z\|_F^2 / \|Z\|_F^2$, $\Delta \Theta = \|\hT - \Theta\|_F^2 / \|\Theta\|_F^2$, and $\Delta P = \|\hP - P\|_F^2/\|P\|_F^2$, for $K$, $Z$, $\Theta$, and $P$, respectively.

\begin{figure}[ht]
    \centering
    \includegraphics[width=1\linewidth]{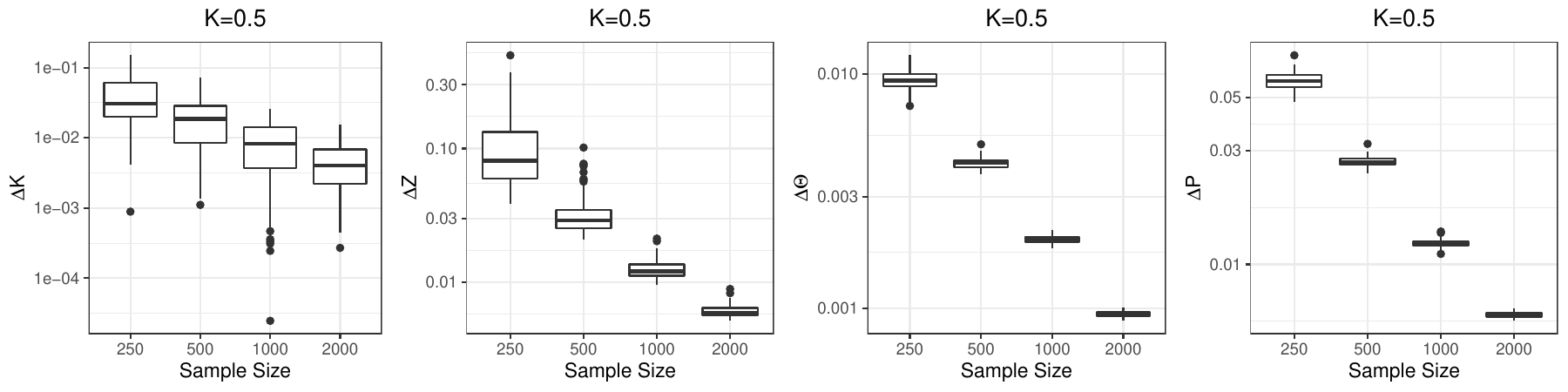}
    \includegraphics[width=1\linewidth]{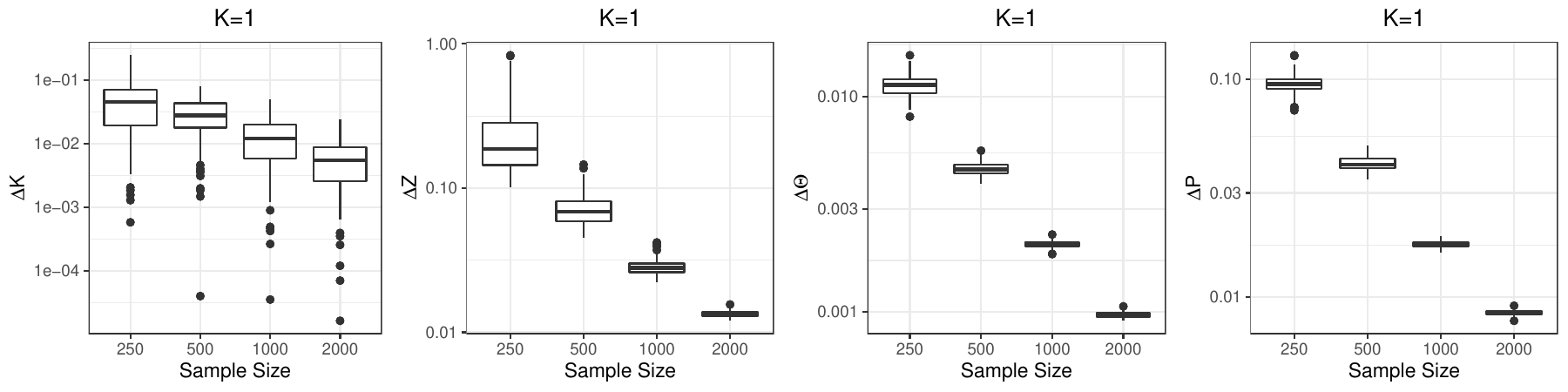}
    \includegraphics[width=1\linewidth]{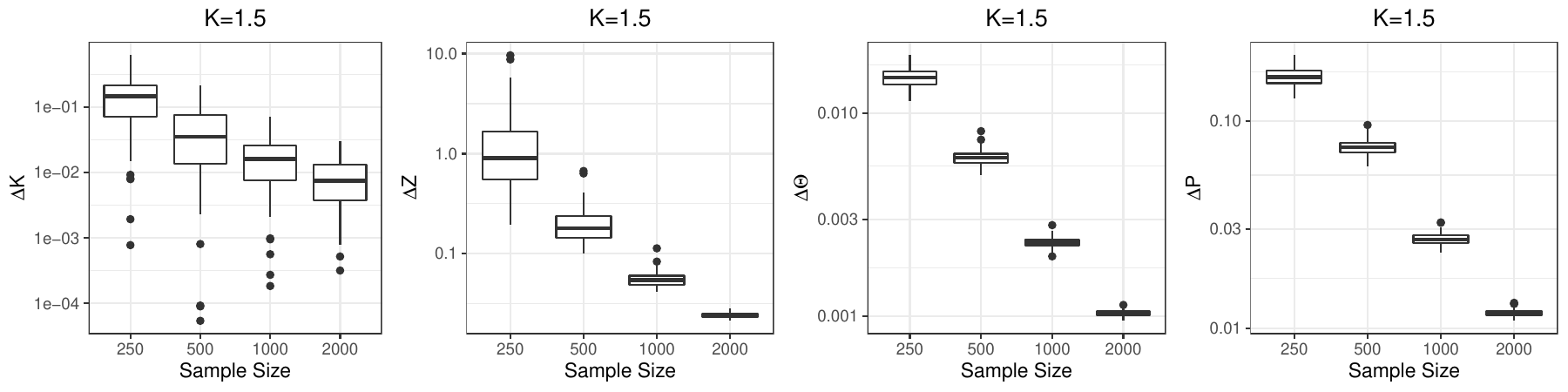}
    \caption{Log-scaled relative errors of maximum-likelihood estimators.}
    \label{fig:consistency}
\end{figure}

\Cref{fig:consistency} shows the box-plots of estimation errors with $100$ replicates under each setting, where the $y$-axes are log-scaled. As the sample size grows larger, we can see clear consistency patterns for all parameter estimates. The estimation errors on $K$ have a relatively larger variance compared with the other three quantities, possibly due to numerical instability issues as the updating formula takes the summation of all pairs of nodes.


\begin{figure}[ht]
    \centering
    \includegraphics[width=\linewidth]{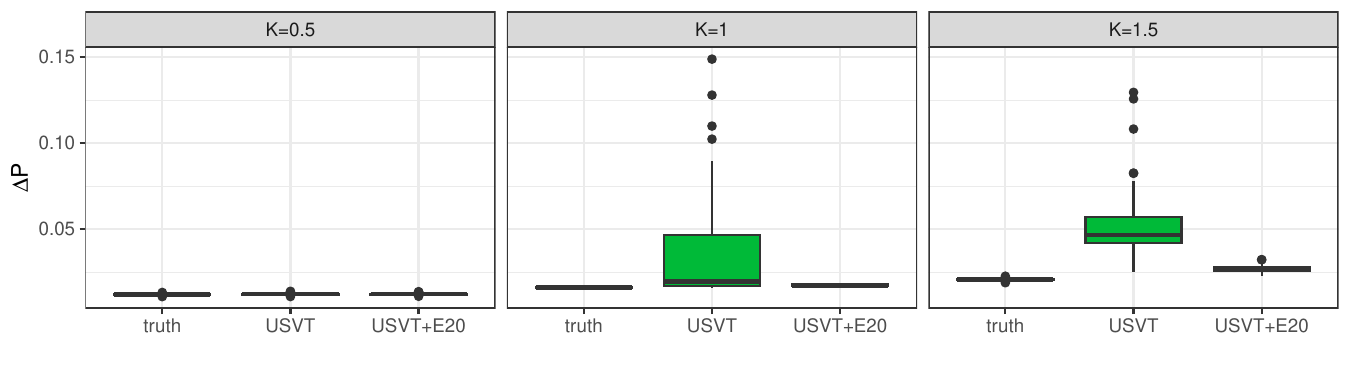}
    \caption{Relative errors of $P$ among two initialization methods and the oracle case where parameters are initialized with true values.}
    \label{fig:compare_initialization}
\end{figure}

We further compare the performances of initialization methods proposed in \Cref{sec:estimation_and_initialization}, where we denote USVT as the initialization method described in \Cref{alg:init_w_USVT} and USVT+E20 as the initialization method combining \Cref{alg:init_w_USVT} with pre-fitting a $20$-dimensional Euclidean model. \Cref{fig:compare_initialization} shows the relative errors of $P$ versus the oracle case where both $K$ and $Z$ are initialized with the true values. We can see that USVT+E20 performs reasonably well for all choices of $K$, whereas USVT has worse performance when the latent space is highly curved due to its limitations as a linear approximation method. It is observed that using USVT followed by a high-dimensional Euclidean model is a more stable and preferred way to initialize \Cref{alg:mgd_gd}. 

\subsection{Embedding Error versus Latent Space Curvature}
\label{sec:simulation_estimation_error_versus_curvature}

\begin{figure}[ht]
    \centering
    \includegraphics[width=\linewidth]{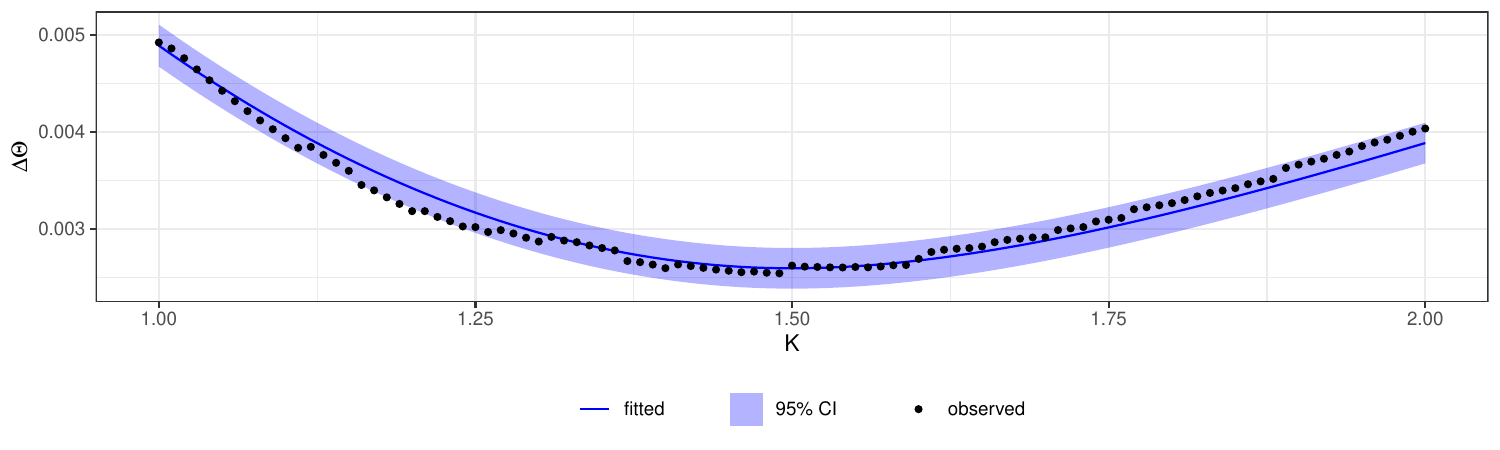}
    \caption{Observed embedding errors from simulations, versus the fitted curve and $95\%$ prediction band by regression on the theoretical bound $f(K) = (1-\min\{\sqrt{K/1.5}, \sqrt{1.5/K} \})^2$.}
    \label{fig:simulation_err_theta}
\end{figure}

We conduct two experiments to examine the estimation error caused by misspecifying the latent space curvature. In the first experiment, we aim to examine the lower bound on the embedding error, $\Delta \Theta$, derived in \Cref{thm:embedding_error}. We simulate a network with $1000$ nodes from the hyperbolic model with $K=1.5$ and fit hyperbolic models with fixed curvature parameters ranging from $1.0$ to $2.0$. Then we compare the observed values of $\Delta \Theta$ from simulations with theoretical values from function $f(K) = (1-\min\{\sqrt{K/1.5}, \sqrt{1.5/K} \})^2$, by fitting a simple linear regression model on the observed $\Delta \Theta$ over values of $f(K)$ for $K \in [1, 2]$. In \Cref{fig:simulation_err_theta} we show the fitted curve and the $95\%$ prediction band, which covers most of the observed values. The observed and fitted values are especially close when $K$ is fixed near the truth of $1.5$. This observation indicates that the lower bound derived in \Cref{thm:embedding_error} could be optimal in terms of the relative error of $K$ when $\hat K$ lies in a small neighborhood of $K$.

\begin{figure}[ht]
    \centering
    \includegraphics[width=\linewidth]{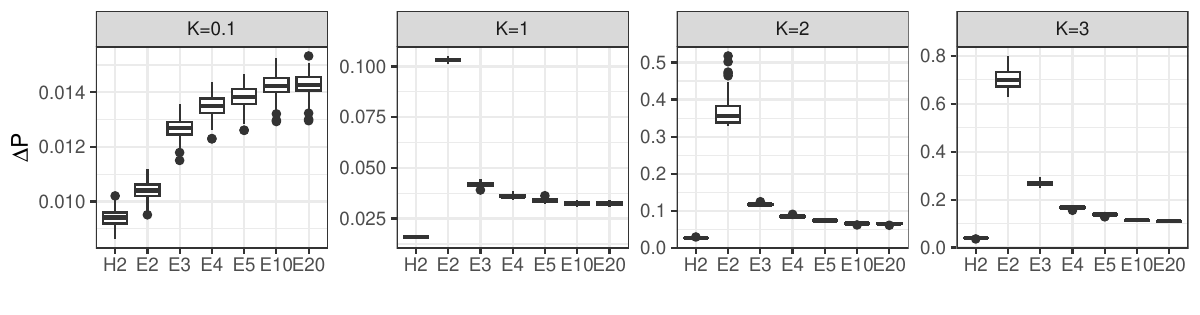}
    \caption{Relative error of the probability matrix, estimated by the hyperbolic model and Euclidean models with different latent space dimensions. ``H2'' represents the two-dimensional hyperbolic model and ``E${x}$'' represents the ${x}$-dimensional Euclidean model.}
    \label{fig:simulation_err_P_EU}
\end{figure}

In the second experiment, we generate networks with $1000$ nodes from the hyperbolic model while using Euclidean models to fit them, with increasing latent space dimensions from $2$ to $20$. We conduct experiments with increasing curvature parameters, from $K=0.1$, where the latent space is closer to Euclidean, to $K = 3$, where the latent space is significantly hyperbolic. \Cref{fig:simulation_err_P_EU} reports the relative error of the estimated probability matrix of different models, averaged over $100$ replicates. In all settings, the hyperbolic model achieves the lowest estimation error since it has the learnable curvature parameter. When $K=0.1$, the performance of $E2$ is comparable to that of $H2$, whereas Euclidean models with higher dimensions show larger estimation errors attributed to overfitting. In the cases of $K=1, 2$, and $3$, $E2$ experiences a considerably high estimation error. Moreover, other Euclidean models with greater latent space dimensions exhibit nearly twice the error of the hyperbolic model. These results demonstrate that the influence of latent space curvature on model fitting differs from, and cannot be substituted by, the impact of latent space dimension, and these two effects are not interchangeable (see Section 5.5).

  \subsection{Statistical Inference of the Curvature Parameter $K$}

  The curvature parameter $K$ characterizes how heterogeneous a network is, which can be reflected from its global properties as shown in Section 5.1. 
  An interesting practical question is how to conduct statistical inference over the curvature parameter. We propose a two-step procedure: (1) conduct a likelihood-ratio test to determine whether the latent space geometry is Euclidean ($K=0$) or hyperbolic ($K >0$); and (2) construct a confidence interval for $K$ if the testing result indicates $K>0$. Given the challenging nature of deriving the asymptotic distribution of the likelihood ratio test statistic in the context of the complex nonlinear latent space model, we employ a parametric bootstrap method to conduct inference in practice. Due to space limitations, the procedure and simulation results are presented in the Supplementary Materials. 

\subsection{Impact of Latent Space Dimension on Model Estimation}
{\color{black}
We also investigate the impact of mis-specifying the latent space dimension on model
estimation through simulation studies, finding that (1) under-specifying latent space dimension is more detrimental to model fitting than over-specifying it; and (2) under-specifying latent space dimension would lead to over-estimating $K$, and vice versa. These results further enhance our understanding of the interplay between latent space curvature and dimension. Detailed results are provided in the Supplementary Materials (Section B.4).
}

\section{Real-world Data Example}
\label{sec:real_data}


In this section, we employ the proposed hyperbolic latent space model to analyze the Simmons College Facebook friendship network \citep{traud2012social}. Within this anonymized network, individual nodes represent students, and edges connect nodes when the corresponding students are mutual friends on Facebook. The dataset offers supplementary details about each student, including their graduation year. To align with \cite{ma2020universal}, we preprocess the dataset, focusing solely on the largest connected component of students who graduated between 2006 and 2009. Subsequent to preprocessing, the network encompasses $1158$ nodes and $24449$ undirected edges.

For visualization purposes, we begin by fitting the $2$-dimensional latent space models. The estimated curvature parameter is $\hK = 1.450$, along with a log-likelihood value of $-149129.6$. The outcome of the testing procedure leads to the rejection of the null hypothesis that the underlying latent space geometry is Euclidean at a $5\%$ significance level. Moreover, the $95\%$ confidence interval for the latent space curvature is $(-1.532, -1.352)$. For comparison, we also fit a $2$-dimensional hyperbolic model with a standard curvature parameter of $K=1$, as well as a $2$-dimensional Euclidean model with a curvature of $0$. The log-likelihoods of these two models are $-157616.8$ and $-166412.0$, respectively. This indicates two key observations: (1) hyperbolic latent space models provide a better fit for the network than the Euclidean model; and (2) selecting an appropriate hyperbolic space curvature can further enhance model fitting. Notably, in terms of parameter count, the $2$-dimensional hyperbolic models only introduce one additional curvature parameter when compared to their $2$-dimensional Euclidean counterpart.

\begin{figure}[ht]
    \centering
    \includegraphics[width=\linewidth]{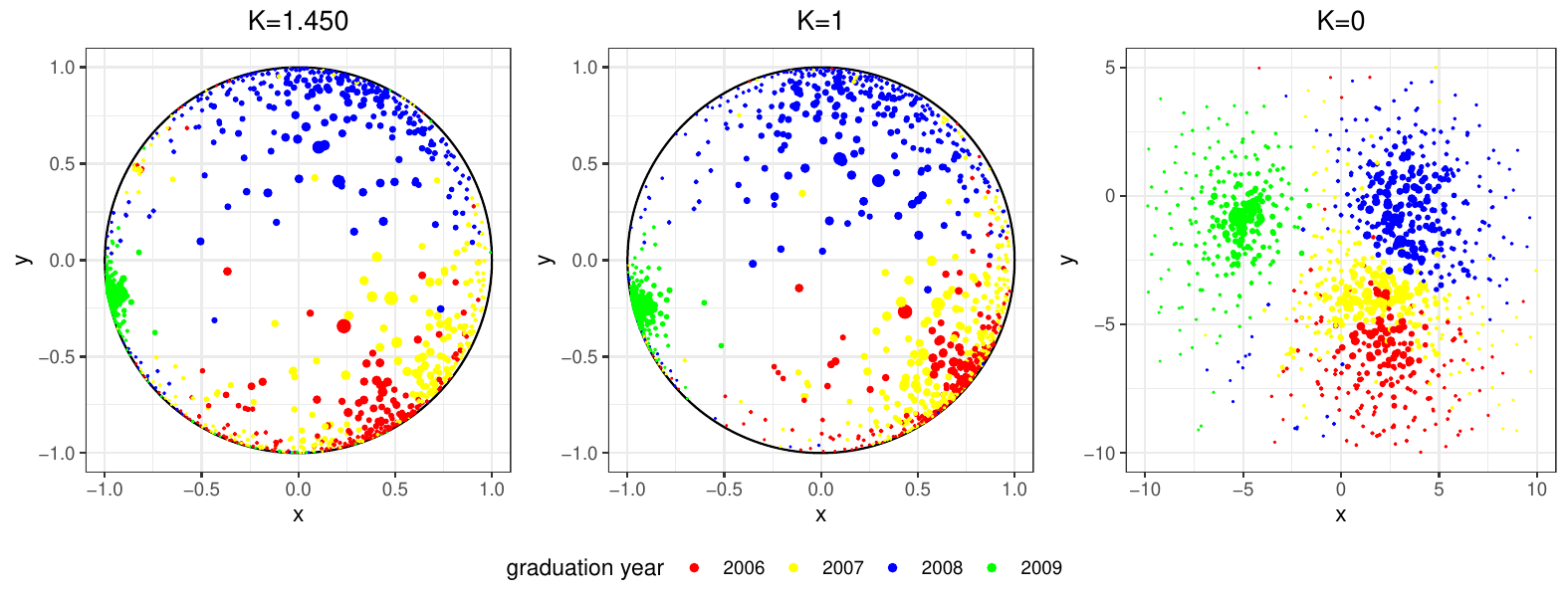}
    \caption{Visualization of estimated latent positions on three latent spaces. Left: $\mathbb{H}_2^{-1.450}$; Middle: $\mathbb{H}_2^{-1}$; Right: $\mathbb{E}_2$. Node color represents graduation year and node size is proportional to node degree.}
    \label{fig:Simmons_Vis}
\end{figure}

The estimated latent positions are illustrated in \Cref{fig:Simmons_Vis}, where node colors are associated with the student's graduation year and node sizes are proportional to the node degree. 
Latent positions of high-degree nodes in hyperbolic spaces are close to the center of the disc, whereas most nodes with small degrees are estimated near the boundary. From \Cref{fig:Simmons_Vis} we can observe a clear clustering pattern among nodes with different graduation years. However, it is noticeable that the nodes from the graduation years 2006 (represented in red) and 2007 (represented in yellow) exhibit a high degree of overlap in the estimated latent positions for both $\mathbb{E}2$ and $\mathbb{H}{2}^{-1}$. On the other hand, by estimating the optimal latent space curvature, the learned latent positions from $\mathbb{H}{2}^{-1.450}$ has the best separation of 2006 and 2007 groups.


To further investigate the impact of the latent space dimension and curvature, we compare network latent space models with different latent dimensions and curvatures with (1) classic model selection criteria BIC and AIC, calculated with the sample size of $\binom{1158}{2}$; and (2) average link prediction AUC scores, evaluated by randomly choosing $80\%$ entries of $A$ as training set and evaluate AUC in the rest of $20\%$ entries, repeated $100$ times. The results of the experiment in Table \ref{tab:auc} indicate that, within each dimension, the proposed hyperbolic model with a learnable curvature exceeds the performance of both the hyperbolic model with fixed $K=1$ and the Euclidean model. For estimating the dimension of the latent space,  BIC prefers the $3$-dimensional model, while AIC is less conservative and favors the $5$-dimensional model. The highest average link prediction AUC score is achieved by the $6$-dimensional model. In the Supplementary Materials, we further compare with the link prediction performance of the Poincar\'e-embedding method proposed in \cite{nickel2017poincare} and demonstrate the advantages of the proposed model, particularly in fitting the optimal curvature and employing robust initialization.

We also note the following observations: (1) Hyperbolic models are particularly powerful in low-dimensional settings. With an optimized curvature, a $3$-dimensional hyperbolic model can achieve similar link prediction performance of a $5$-dimensional Euclidean model. (2) The estimation of $K$ becomes smaller as the latent space dimension of the working model grows and the hyperbolic models have less improvement than the Euclidean model. 
Intuitively, this may be attributed to the trade-off between the latent space dimension and the curvature in achieving optimal model capacity. Increasing the dimension enhances the model's capacity, while making the space more hyperbolic improves its ability to capture the network properties. 
Such intuition also explains why hyperbolic models with fixed $K=-1$ is worse than Euclidean models for $d \geq 5$. A similar phenomenon can be found in \cite{mathieu2019continuous}. (3) Nevertheless, for each $2 \leq d \leq 7$, the proposed hyperbolic model with learnable curvature still outperforms the other two models for higher-dimensional cases for all four criteria reported in Table \ref{tab:auc}. And for all latent space dimensions, the bootstrap tests reject the null hypothesis of $K=0$. 
These observations shed light on  the significance of accounting for the latent space curvature in network analysis.

\begin{table}[]
\centering
\begin{tabular}{|c|l|c|c|c|c|}
\hline
Dimension              & \multicolumn{1}{c|}{Model} & \multicolumn{1}{c|}{Log-Likelihood ($\times 10^5$)} & \multicolumn{1}{c|}{BIC ($\times 10^5$)} & \multicolumn{1}{c|}{AIC ($\times 10^5$)} & \multicolumn{1}{c|}{AUC (sd)} \\ \hline
\multirow{3}{*}{d = 2} & $\mathbb{E}_2$             & -1.664                            & 3.639                      & 3.375                      & 0.855 (0.004)                \\
                       & $\mathbb{H}_2^{-1}$        & -1.577                            & 3.463                      & 3.199                      & 0.870 (0.002)                \\
                       & $\mathbb{H}_2^{-1.450}$    & -1.491                            & 3.293                      & 3.029                      & 0.884 (0.002)                \\ \hline
\multirow{3}{*}{d = 3} & $\mathbb{E}_3$             & -1.425                            & 3.315                      & 2.919                      & 0.883 (0.002)                \\
                       & $\mathbb{H}_3^{-1}$        & -1.358                            & 3.183                      & 2.787                      & 0.903 (0.002)                \\
                       & $\mathbb{H}_3^{-1.283}$    & -1.356                            & \textbf{3.179}             & 2.782                      & 0.902 (0.002)                \\ \hline
\multirow{3}{*}{d = 4} & $\mathbb{E}_4$             & -1.343                            & 3.308                      & 2.779                      & 0.895 (0.002)                \\
                       & $\mathbb{H}_4^{-1}$        & -1.318                            & 3.257                      & 2.728                      & 0.908 (0.002)                \\
                       & $\mathbb{H}_4^{-0.516}$    & -1.304                            & 3.230                      & 2.702                      & 0.909 (0.002)                \\ \hline
\multirow{3}{*}{d = 5} & $\mathbb{E}_5$             & -1.306                            & 3.389                      & 2.728                      & 0.903 (0.002)                \\
                       & $\mathbb{H}_5^{-1}$        & -1.310                            & 3.397                      & 2.736                      & 0.910 (0.002)                \\
                       & $\mathbb{H}_5^{-0.272}$    & -1.285                            & 3.349                      & \textbf{2.687}             & 0.911 (0.002)                \\ \hline
\multirow{3}{*}{d = 6} & $\mathbb{E}_6$             & -1.291                            & 3.513                      & 2.720                      & 0.907 (0.002)                \\
                       & $\mathbb{H}_6^{-1}$        & -1.306                            & 3.544                      & 2.751                      & 0.911 (0.002)                \\
                       & $\mathbb{H}_6^{-0.170}$    & -1.275                            & 3.483                      & 2.689                      & \textbf{0.912} (0.002)       \\ \hline
\multirow{3}{*}{d = 7} & $\mathbb{E}_7$             & -1.279                            & 3.645                      & 2.720                      & 0.909 (0.002)                \\
                       & $\mathbb{H}_7^{-1}$        & -1.303                            & 3.694                      & 2.768                      & 0.911 (0.002)                \\
                       & $\mathbb{H}_7^{-0.121}$    & -1.269                            & 3.625                      & 2.699                      & 0.912 (0.002)                \\ \hline
\end{tabular}
\caption{Log-likelihoods, BIC, AIC, and average link prediction AUC scores of latent space models with different dimensions and curvatures. Best BIC, AIC, and AUC values are highlighted in bold.}
\label{tab:auc}
\end{table}

In conclusion, our data analysis shows the significant advantages of the proposed hyperbolic latent space model. It not only offers better model fitting and visualization capabilities but also enhances downstream task performance, demonstrating its potential as a valuable tool for analyzing and understanding complex network data.


\section{Discussion}
\label{sec:discussion}

This work proposes a hyperbolic network latent space model with a learnable curvature parameter to capture the effect of latent space geometry. The importance of latent space curvature is demonstrated with theoretical results combined with experiments on simulated and real-world network data. In particular, \Cref{thm:embedding_error} shows that a learnable curvature parameter is essential for all hyperbolic embedding methods beyond network latent space models. \Cref{thm:consistency_theta} and \Cref{thm:consistency_KZ} establish the consistency rates for maximum-likelihood estimators. Through simulation experiments, we study the geometric effect of curvature on model properties, stability of estimation procedure, and embedding errors. We demonstrate that estimating the latent space curvature benefits model fitting and downstream task performance on the Simmons Facebook friendship network.

Our work also raises several promising directions for future research. From a modeling perspective, one may include additional node and edge parameters in \Cref{eq:model} to further incorporate node-level or edge-wise information. For example, similar to \cite{ma2020universal} we can let $P_{ij} = \sigma(d_{\mathbb{H}_d^{-K}}(z_i, z_j) + \alpha_i + \alpha_j + \beta x_{ij})$, with $\alpha_i, 1 \leq i \leq n$ accounting for node heterogeneity and $x_{ij}, 1 \leq i < j \leq n$ representing edge-related covariates. The node parameters $\alpha_i$ could be unbounded to better model the sparse network data. These extensions that incorporate more realistic settings are worthwhile for future exploration.
Another interesting generalization of the proposed model is to consider latent spaces with non-constant local curvatures. Taking social networks as an example, the local curvature near the latent positions of hub nodes might be 
larger such that the pairwise distances on the latent space are shorter among these nodes. It is also interesting to further investigate the model selection problem on the interplay between latent space curvature and the dimension, which relates to a technical problem of generalizing \Cref{thm:embedding_error} towards higher-dimensional hyperbolic spaces. 

Beyond network data, hyperbolic latent space models or embedding methods can be used for modeling tree structured data (e.g., phylogenetic tree; \citealp{sala2018representation}) and other heterogeneous relational data (e.g., text data parsed as graphs; \citealp{tifrea2018poincar}). With the insights provided by \Cref{thm:embedding_error}, it would be interesting to investigate whether including the learnable curvature parameter can help increase the downstream task performance of these general hyperbolic embedding methods.


\spacingset{1.2}
 \small 
\bibliography{ref}

\newpage
\spacingset{1.9}
 \small 
\appendix

\end{document}